\documentclass[aps,prd,twocolumn,superscriptaddress,nofootinbib,floatfix]{revtex4-2}
\usepackage{amsmath,amssymb,graphicx,bm,booktabs,placeins}
\allowdisplaybreaks
\usepackage[colorlinks=true,linkcolor=blue,citecolor=blue,urlcolor=blue]{hyperref}

\newcommand{\qqbar}{\langle\bar qq\rangle}
\newcommand{\GG}{\langle g_s^2G^2\rangle}
\newcommand{\mix}{\langle\bar q g_s\sigma\!\cdot\! G q\rangle}

\begin{document}
\raggedbottom

\title{Magnetic moments of the $\Sigma_c$ and $\Sigma_b$ baryons from external-field QCD sum rules}

\author{Halil Mutuk}
\email{hmutuk@omu.edu.tr}
\affiliation{Department of Physics, Faculty of Sciences, Ondokuz Mayis University, 55139 Samsun, T\"urkiye}

\date{\today}

\begin{abstract}
We compute the magnetic moments of the $\Sigma_c$ and $\Sigma_b$ isotriplets with external-field QCD sum rules.
The heavy-quark propagator is kept with its full mass dependence and with its couplings to the photon, to soft
gluons and to both simultaneously, and the magnetic sum rule is normalized by the mass sum rule of the same
interpolating field, which removes the residue and gives the baryon masses within a few percent. The quark-condensate magnetic susceptibility $\chi$, which
otherwise dominates the uncertainty, is fixed by the proton magnetic moment computed with the same interpolating
field; the neutron moment then follows to about $5\%$.
Uncertainties from the QCD inputs are evaluated with a Monte Carlo analysis, and those from the interpolating field,
the Borel window, the scale of the heavy-quark mass, the calibration and the normalization are added.
With $\overline{\rm MS}$ heavy-quark masses we obtain
$\mu_{\Sigma_c^{++}}=2.37\pm0.65$, $\mu_{\Sigma_c^{+}}=0.50\pm0.15$, $\mu_{\Sigma_c^{0}}=-1.39\pm0.37$,
$\mu_{\Sigma_b^{+}}=2.16\pm0.65$, $\mu_{\Sigma_b^{0}}=0.55\pm0.16$ and $\mu_{\Sigma_b^{-}}=-1.06\pm0.32$ in nuclear
magnetons; at this order the results depend sizably on the heavy-quark mass scheme. The moments are carried by the
light diquark; the heavy-quark contribution, whose separation depends on the interpolating field, is
$-0.13\pm0.02\,\mu_N$ for $\Sigma_c$ and $+0.013\pm0.003\,\mu_N$ for $\Sigma_b$.
\end{abstract}

\maketitle

\section{Introduction}

Magnetic moments of singly heavy baryons probe how the magnetic response is shared between a heavy quark and a light
diquark. In the heavy-quark limit the light-diquark spin in $\Sigma_Q$ decouples from the heavy-quark spin; the
light-diquark contribution to the moment is then the same for $\Sigma_c$ and $\Sigma_b$, while the heavy-quark
contribution is suppressed by $e_Q/m_Q$. None of these moments has been measured. Measurements of charm-baryon
magnetic moments are, however, being prepared. In a fixed-target experiment at the LHC, charm baryons produced in
a target are channeled in a bent crystal, whose strong interplanar electric field induces a measurable spin
precession~\cite{Baryshevsky:2016,Fomin:2017,Bagli:2017,Fomin:2020,Aiola:2021}. The dedicated experiment ALADDIN,
proposed in a Letter of Intent to the LHCC~\cite{ALADDIN:2024}, aims to measure the moments of $\Lambda_c^+$ and
$\Xi_c^+$, whose lifetimes are $2$--$4\times10^{-13}$~s, with projected uncertainties of $0.012$ and $0.018\,\mu_N$
for the baseline crystal configuration after two years of data taking in LHC Run~4. The method requires weakly decaying baryons, so that the strongly
decaying $\Sigma_Q$ states are not accessible directly. In the constituent quark model, however, the $\Lambda_c^+$
moment equals the charm-quark moment~\cite{Fomin:2020,ALADDIN:2024}, which also enters the $\Sigma_c$ moments; such
measurements will therefore constrain the heavy-quark contribution and test the approaches used for heavy-baryon
moments, including their predictions for $\Sigma_Q$. Predictions exist from quark
models~\cite{JuliaDiaz:2004,Faessler:2006,Bernotas:2012,Sharma:2010,Barik:1983,Kumar:2005,Patel:2008,Simonis:2018,Mohan:2022,Dhir:2013,Majethiya:2008}, lattice QCD
for $\Sigma_c$~\cite{Can:2014}, chiral perturbation theory~\cite{Wang:2018,Shi:2019}, the chiral quark-soliton
model~\cite{Yang:2018}, external-field QCD sum rules~\cite{Zhu:1997} and, recently, light-cone QCD sum
rules~\cite{Ozdem:2025}; for $\Sigma_c^{++}$ they range from $1.3$ to
$3.1\,\mu_N$. Light-cone QCD sum rules have also been applied to the magnetic moments of $\Xi_Q$
baryons~\cite{Aliev:2008xi}, of spin-$3/2$ heavy baryons~\cite{Aliev:2009sp} and to the $\Sigma_Q\to\Lambda_Q$
transitions~\cite{Aliev:2002tr}.

In the external-field method~\cite{Ioffe:1984,Balitsky:1983} the baryon two-point correlator is evaluated in a weak
static electromagnetic field, which couples to the quark propagators and, through vacuum susceptibilities, to the QCD
vacuum. The only external-field calculation for heavy baryons, Ref.~\cite{Zhu:1997}, treated the $\Sigma_c$ triplet
with the Ioffe-type current. It kept the perturbative term and a subset of condensate contributions, did not include
the gluon condensate or the couplings of the charm quark to gluons, took the residue $\lambda^2$ from independent mass
sum rule analyses, included leading-order anomalous-dimension factors, and estimated the uncertainty by varying the
susceptibility $\chi$. Light-cone sum rules~\cite{Ozdem:2025} describe the long-distance photon emission through
photon distribution amplitudes, whereas in the external-field method it enters through the vacuum susceptibilities;
the two approaches are complementary, with different nonperturbative inputs and different sources of uncertainty.

In this work we
(i)~keep the full heavy-quark mass dependence in all terms, including the couplings of the heavy quark to the photon,
to soft gluons, to both simultaneously, and to the gluon condensate;
(ii)~normalize the magnetic sum rule by the mass sum rule of the same current and the same OPE, so that $\lambda^2$
cancels and the mass sum rule serves as a check;
(iii)~treat both the $\Sigma_c$ and the $\Sigma_b$ triplets;
(iv)~fix $\chi$ with the proton magnetic moment computed with the same current, and evaluate the uncertainties
with a Monte Carlo analysis~\cite{Leinweber:1997,Wang:2008oct} together with the dependence on the interpolating
field, the Borel window and the heavy-quark mass scheme; and
(v)~separate the light- and heavy-quark contributions.
This is, to our knowledge, the first external-field analysis of the $\Sigma_b$ moments, with the light- and
heavy-quark contributions separated and with an uncertainty budget that includes the condensates and susceptibilities, the interpolating field,
the Borel window and the scale of the heavy-quark mass. The new OPE ingredients (gluon-condensate terms and the
simultaneous coupling of the heavy quark to the photon and to gluons) turn out to be numerically small, about $1\%$ of
the $W_{E1}$ OPE; the main differences with respect to Ref.~\cite{Zhu:1997} come from the normalization, the treatment
of $\chi$ and the error analysis. Unlike Ref.~\cite{Zhu:1997}, we do not include anomalous-dimension factors.

\section{Sum rules}

For $\Sigma_Q=(qqQ)$ with $q=u$ or $d$ we use
\begin{equation}
\eta=-2\epsilon^{abc}\left[(q_a^TC\gamma_5Q_b)q_c+\beta\,(q_a^TCQ_b)\gamma_5q_c\right],
\label{eq:current}
\end{equation}
where $C$ is the charge-conjugation matrix and $a,b,c$ are color indices. This is the proton field of
Refs.~\cite{Ioffe:1981,Wang:2008oct} with $d\to Q$. For $\beta=-1$ (Ioffe current) it is, after a Fierz
transformation, the standard $\Sigma_Q$ current with the heavy quark outside a spin-one light diquark.
The $I_3=0$ states $(udQ)$ are obtained from isospin symmetry,
$\mu_{\Sigma_Q^{I_3=0}}=\frac12\left(\mu_{\Sigma_Q^{I_3=+1}}+\mu_{\Sigma_Q^{I_3=-1}}\right)$; this avoids the
admixture of $\Lambda_Q$ that a current of the form~(\ref{eq:current}) would have for two different light flavors.

We consider $\Pi(p)=i\int d^4x\,e^{ip\cdot x}\langle0|T\{\eta(x)\bar\eta(0)\}|0\rangle_F$ in a static field $F_{\mu\nu}$
and the structure $W_{E1}=F_{\mu\nu}(\hat p\sigma^{\mu\nu}+\sigma^{\mu\nu}\hat p)$, $\hat p=\gamma\cdot p$, which is
chiral-even in the light-quark sector and was found to be the most reliable one for the nucleon~\cite{Wang:2008oct}.
After Borel transformation the phenomenological sides of the $W_{E1}$
and of the $\hat p$ mass sum rules are
\begin{align}
\Pi_{E1}(M^2)&=-\frac{\tilde\lambda^2}{4}\left(\frac{\mu}{M^2}+A\right)e^{-m_B^2/M^2},\nonumber\\
\Pi_{\hat p}(M^2)&=\tilde\lambda^2e^{-m_B^2/M^2},
\label{eq:phen}
\end{align}
where $\mu$ is the magnetic moment in natural magnetons and $A$ accounts for transitions to excited states, which are
not suppressed by the Borel transformation. Their ratio,
\begin{equation}
R(M^2)\equiv-4\,\frac{\Pi_{E1}(M^2)}{\Pi_{\hat p}(M^2)}=\frac{\mu}{M^2}+A ,
\label{eq:ratio}
\end{equation}
is free of $\tilde\lambda^2$ and $m_B$; $\mu\,m_N/m_B$ is the moment in nuclear magnetons.

\emph{OPE.} For the light quarks ($m_q=0$) we use the propagator in the Fock--Schwinger gauge to first order in
$F_{\mu\nu}$ and up to order $x^4$~\cite{Ioffe:1984,Balitsky:1983,Wang:2008oct},
\begin{align}
S_q^{ab}(x)&=\Big[\frac{i\hat x}{2\pi^2x^4}-\frac{\qqbar}{12}+\frac{x^2}{192}\mix\nonumber\\
&\qquad-\frac{x^4}{3^32^{10}}\qqbar\GG\Big]\delta^{ab}\nonumber\\
&+\Big[\frac{i}{32\pi^2x^2}g_sG^n_{\alpha\beta}(\hat x\sigma^{\alpha\beta}+\sigma^{\alpha\beta}\hat x)
-\frac{\mix}{192}\sigma_{\alpha\beta}\nonumber\\
&\qquad+\frac{x^2}{3^22^{10}}\qqbar\GG\sigma_{\alpha\beta}\Big]\Big(\frac{\lambda^n}{2}\Big)^{ab}\nonumber\\
&+\frac{ie_q}{32\pi^2x^2}F_{\alpha\beta}(\hat x\sigma^{\alpha\beta}+\sigma^{\alpha\beta}\hat x)\delta^{ab}\nonumber\\
&-\frac{e_q}{24}\chi\qqbar F_{\alpha\beta}\sigma^{\alpha\beta}\delta^{ab}\nonumber\\
&+\frac{e_q\qqbar}{288}F_{\alpha\beta}\big(x^2\sigma^{\alpha\beta}-2x_\rho x^\beta\sigma^{\rho\alpha}\big)\delta^{ab}\nonumber\\
&+\frac{e_q\qqbar}{576}F_{\alpha\beta}\big[x^2(\kappa+\xi)\sigma^{\alpha\beta}\nonumber\\
&\qquad-(2\kappa-\xi)x_\rho x^\beta\sigma^{\rho\alpha}\big]\delta^{ab}\nonumber\\
&-\frac{e_q\qqbar}{16}\Big(\kappa F_{\alpha\beta}-\frac i4\xi\epsilon_{\alpha\beta\mu\nu}F^{\mu\nu}\gamma_5\Big)\Big(\frac{\lambda^n}{2}\Big)^{ab},
\label{eq:Sq}
\end{align}
where the terms with open indices $\alpha\beta$ and $(\lambda^n/2)^{ab}$ are contracted with the gluon field
$g_sG^n_{\alpha\beta}$ emitted by another quark line, and the susceptibilities are defined by
$\langle\bar q\sigma_{\mu\nu}q\rangle_F=e_q\chi\qqbar F_{\mu\nu}$,
$\langle\bar qg_sG_{\mu\nu}q\rangle_F=e_q\kappa\qqbar F_{\mu\nu}$ and
$\langle\bar qg_s\epsilon_{\mu\nu\rho\lambda}G^{\rho\lambda}\gamma_5q\rangle_F=ie_q\xi\qqbar F_{\mu\nu}$;
$\chi$ describes the tensor quark condensate induced by the external field.

For the heavy quark we work in momentum space. In the Fock--Schwinger gauge the photon and gluon fields combine into a
constant field strength $\mathcal F_{\mu\nu}=e_QF_{\mu\nu}+g_sG_{\mu\nu}$, and the propagator is generated recursively,
\begin{align}
S_{n+1}(k)&=\tfrac12\,S_0(k)\,\gamma^\alpha\mathcal F_{\alpha\beta}\,\frac{\partial}{\partial k_\beta}S_n(k),\nonumber\\
S_0(k)&=\frac{i(\hat k+m_Q)}{k^2-m_Q^2},
\label{eq:recursion}
\end{align}
with the same coupling convention as in Eq.~(\ref{eq:Sq}). To the order needed here,
\begin{align}
S_Q(k)&=\frac{i(\hat k+m_Q)}{k^2-m_Q^2}\nonumber\\
&+\frac{i}{4}\,\frac{\sigma_{\alpha\beta}(\hat k+m_Q)+(\hat k+m_Q)\sigma_{\alpha\beta}}{(k^2-m_Q^2)^2}\nonumber\\
&\qquad\times\big(e_QF^{\alpha\beta}+g_sG^{\alpha\beta}\big)\nonumber\\
&+\frac{i}{12}\,m_Q\GG\,\frac{k^2+m_Q\hat k}{(k^2-m_Q^2)^4}\nonumber\\
&+S_Q^{(FG)}(k)+S_Q^{(F\langle G^2\rangle)}(k),
\label{eq:SQ}
\end{align}
with $g_sG^{\alpha\beta}=g_sG^{n\,\alpha\beta}\lambda^n/2$ (contracted with a gluon from a light-quark line). The
second term is the first-order term $S_1$; for $m_Q\to0$ its Fourier transform coincides with the corresponding
term of Eq.~(\ref{eq:Sq}). The third term follows from the second order in $g_sG$ averaged over the gluon condensate,
$\langle g_s^2G^a_{\alpha\beta}G^b_{\mu\nu}\rangle=\frac{\delta^{ab}}{96}\GG(g_{\alpha\mu}g_{\beta\nu}-g_{\alpha\nu}g_{\beta\mu})$,
consistent with $\langle\bar QQ\rangle=-\GG/(48\pi^2m_Q)$. $S_Q^{(FG)}$ is the mixed term of order $e_QF\,g_sG$ (second
order of Eq.~(\ref{eq:recursion})), and $S_Q^{(F\langle G^2\rangle)}$ the term of order $e_QF(g_sG)^2$ averaged over
$\GG$ (third order). We include all terms up to operator dimension eight,
counting $\qqbar$ and $\chi\qqbar$ as dimension three, $\GG$ as four, and $\mix$, $\kappa\qqbar$ and $\xi\qqbar$ as
five, in the constant-field approximation (i.e.\ without $\langle g_s^3G^3\rangle$ and derivative terms).
$\mathcal O(\alpha_s)$ and anomalous-dimension corrections are not included; the anomalous dimensions of the heavy-quark
current and of the operators differ from the light-baryon case and would require a separate study.

\emph{Construction.} After the Dirac algebra (FeynCalc~\cite{FeynCalc1,FeynCalc9,FeynCalc10}) each term has the form
$L(x)\otimes S_Q(k)$, where $L\propto x^{\mu_1}\cdots x^{\mu_j}(-x^2)^{-\nu}$ collects the light quarks, and
$\Pi(p)=i\int\frac{d^4k}{(2\pi)^4}S_Q(k)\tilde L(p-k)$. The Fourier transform
$\int d^4x\,e^{iq\cdot x}(-x^2)^{-\nu}=-i\pi^2\,4^{2-\nu}\frac{\Gamma(2-\nu)}{\Gamma(\nu)}(-q^2)^{\nu-2}$
and its derivatives give $\tilde L$. The loop integral is done with a Feynman parameter $y$, leading to
$\Delta=y(1-y)(\tilde m^2-p^2)$ with $\tilde m^2=m_Q^2/(1-y)$, and is followed by the Borel transformation
$\mathcal B[\Gamma(-c)(\tilde m^2-p^2)^c]=(M^2)^{c+1}e^{-\tilde m^2/M^2}$.
The light-quark powers are regulated analytically, $\nu\to\nu+\varepsilon$. The $1/\varepsilon$ poles appear only in
end-point contributions at $y\to0$ and cancel against zeros of $1/\Gamma(\nu)$ for light-quark structures that are
polynomial in $x$ (e.g.\ $\qqbar^2$); they leave finite pole terms proportional to $e^{-m_Q^2/M^2}$. No logarithm of a
renormalization scale remains. Every term has its spectral weight at $s=\tilde m^2(y)$: a continuous density
$(s-\tilde m^2)^c$ for $c>-1$, and a $\delta$ function or its derivatives for $c\le-1$. We remove all weight above
$s_0$, i.e.\ the $y$ integration is restricted to $\tilde m^2(y)<s_0$; the pole terms at $s=m_Q^2$ are kept.
The perturbative spectral densities are, with $r=m_Q^2/s$,
\begin{align}
\rho_{\hat p}(s)&=\frac{5+2\beta+5\beta^2}{512\pi^4}\,s^2\nonumber\\[2pt]
&\quad\times\left(1-8r+8r^3-r^4-12r^2\ln r\right),\nonumber\\
\rho_{E1}(s)&=\frac{s}{512\pi^4}\Big[-4(3+2\beta+3\beta^2)e_q(1-r)^3\nonumber\\
&\quad+(1+\beta)^2e_Q\left(2+3r-6r^2+r^3+6r\ln r\right)\Big].
\label{eq:rho}
\end{align}
The perturbative heavy-quark term vanishes for $\beta=-1$; the heavy-quark contribution then comes from the
mass-dependent condensate terms. The complete OPE for $\beta=-1$ is given in Appendix~\ref{app:ope}.

\emph{Validation.} (i)~For the pseudoscalar heavy--light correlator the construction reproduces the known perturbative,
quark-condensate, mixed-condensate and gluon-condensate contributions to ten digits for $m_Q=1.3$ and $4.7$~GeV.
(ii)~For $m_Q\to0$ and $Q\to d$ the free, photon and soft-gluon terms of the heavy line reproduce the corresponding
terms of the proton sum rules obtained with the light-quark propagator~(\ref{eq:Sq}): the mass sum rules at both
chiral structures exactly, and the $W_{E1}$ magnetic sum rule term by term, including the susceptibility terms. The
heavy-quark terms with a gluon condensate or with simultaneous photon and gluon coupling have no counterpart in the
light-quark propagator and are not part of this comparison; in particular, the $e_Q\qqbar\mix/M^4$ term of
Appendix~A, which survives for $m_Q\to0$, comes entirely from the $e_QF\,g_sG$ term. In the all-light case the $W_{E1}$ sum rule has no $m_0^2\qqbar^2$ term:
all diagrams contributing to it are polynomial in $x$ and vanish after Borel transformation.
(iii)~The recursion~(\ref{eq:recursion}) reproduces both the photon vertex and the gluon-condensate term with a single
normalization.

\section{Numerical analysis}

\begin{table}[t]
\caption{Input parameters and the Gaussian widths used in the Monte Carlo analysis. The widths of $\kappa$ and $\xi$
are our choice; $\chi$ is fixed in each Monte Carlo sample by the proton magnetic moment (central value and
calibration uncertainty; see text).}
\label{tab:inputs}
\centering\small\setlength{\tabcolsep}{4pt}
\begin{tabular}{lll}
\toprule
Parameter & Value & Ref.\\
\midrule
$a=-(2\pi)^2\qqbar$ & $0.52\pm0.05\ \text{GeV}^3$ & \cite{Wang:2008oct}\\
$\langle\alpha_sG^2/\pi\rangle$ & $0.012\pm0.004\ \text{GeV}^4$ & \cite{SVZ:1979}\\
$m_0^2=\mix/(-\qqbar)$ & $0.72\pm0.07\ \text{GeV}^2$ & \cite{Wang:2008oct}\\
$\chi$ & $-5.66\pm0.61\ \text{GeV}^{-2}$ & $\mu_p$\\
$\kappa$ & $0.75\pm0.38$ & \cite{Wang:2008oct}\\
$\xi$ & $-1.5\pm0.75$ & \cite{Wang:2008oct}\\
$m_c(m_c)$, $m_b(m_b)$ & $1.2730(46)$, $4.183(7)$ GeV & \cite{PDG:2024}\\
\bottomrule
\end{tabular}
\end{table}

The inputs are listed in Table~\ref{tab:inputs}. The heavy-quark masses are taken in the $\overline{\rm MS}$ scheme
and the baryon masses from Ref.~\cite{PDG:2024}.

\emph{Calibration of $\chi$.} Estimates of $\chi$ range from about $-3$ to
$-9$~GeV$^{-2}$~\cite{Belyaev:1984,Ball:2003,Vainshtein:2003}, partly because $\chi$ depends on the renormalization
scale (the windows correspond to $\sqrt{M^2}\simeq1.4$ and $2.1$~GeV for $\Sigma_c$ and $\Sigma_b$); the lattice gives
$-2.08(8)$~GeV$^{-2}$ at 2~GeV~\cite{Bali:2012}. Since $\chi$ dominates the uncertainty, we fix it with the proton
magnetic moment. The proton sum rules follow from the same current with $Q\to d$ (massless); we evaluate them with
the same OPE truncation and the ratio method and require $\mu_p=2.793\,\mu_N$~\cite{PDG:2024}. Because the
sum rule is linear in $\chi$, this is done in every Monte Carlo sample with that sample's other inputs, so that the
variations common to the nucleon and $\Sigma_Q$ sum rules largely cancel. For the proton the $\qqbar^2$ term is the
leading power correction and the dimension $\geq6$ criterion used below cannot be met; we require instead that
dimension-eight terms contribute at most $10\%$ of the mass OPE and that the pole contribution exceeds $50\%$, which
together with the proton mass gives $M^2=0.8$--$1.1$~GeV$^2$ and $s_0=2.3$--$2.6$~GeV$^2$. We calibrate at
$\beta=-1$, where these criteria hold, and use the same $\chi$ for all values of $\beta$. This yields
$\chi=-5.66$~GeV$^{-2}$ at central inputs; over the Monte Carlo samples, where it follows the other inputs, its
spread is $\pm0.6$~GeV$^{-2}$ (this spread is part of the Monte Carlo uncertainty). As a test, the neutron
sum rule (the same OPE with $e_u\leftrightarrow e_d$) then gives $\mu_n=-1.81\,\mu_N$ and $\mu_n/\mu_p=-0.65$, compared
with $-1.913\,\mu_N$ and $-0.685$, without further adjustment. The neutron moment itself depends strongly on $\chi$
($-1.22$, $-1.54$, $-1.89$ and $-2.36\,\mu_N$ for $\chi=-3.15$, $-4.5$, $-6$ and $-8$~GeV$^{-2}$), whereas the
ratio $\mu_n/\mu_p$ varies only between $-0.76$ and $-0.61$; the test thus confirms mainly the isospin structure of
the sum rule. Since approximations at the $10\%$ level enter the derivation of leading-order baryon sum
rules~\cite{Leinweber:1997}, consistent with the $5\%$ deviation of $\mu_n$, we assign a $10\%$ theoretical
uncertainty to the nucleon sum rule, which corresponds to $\delta\chi=0.60$~GeV$^{-2}$, and add the variation for
proton windows shifted or enlarged by $0.1$~GeV$^2$ within the criteria ($0.14$~GeV$^{-2}$), giving a calibration
uncertainty $\delta\chi=0.61$~GeV$^{-2}$, which we add as a separate systematic uncertainty. The calibrated $\chi$ is an effective parameter of the leading-order sum
rules: it is two to three times larger in magnitude than the lattice value and absorbs part of the corrections common
to the nucleon and $\Sigma_Q$ sum rules. For comparison, $\chi=-6$~GeV$^{-2}$ was used for the octet baryons in
Ref.~\cite{Wang:2008oct}. The windows of the
proton and the $\Sigma_Q$ sum rules lie at different scales, $\sqrt{M^2}\simeq1.0$, $1.4$ and $2.1$~GeV. At leading
order the $\chi\qqbar^2$ term, which carries the $\chi$ dependence, scales as $\alpha_s^{-8/27}$ (from
$\chi\qqbar\propto\alpha_s^{4/27}$ and $\qqbar\propto\alpha_s^{-4/9}$); between these scales it grows by $9\%$ for
$\Sigma_c$ and $16\%$ for $\Sigma_b$. Since anomalous dimensions are not included elsewhere, we do not apply this
factor but add the corresponding shift to the calibration uncertainty.

\begin{figure*}[t]
\centering
\includegraphics[width=0.45\textwidth]{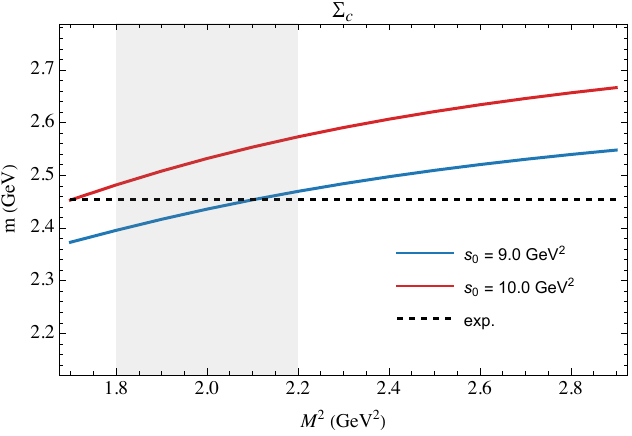}\hfill
\includegraphics[width=0.45\textwidth]{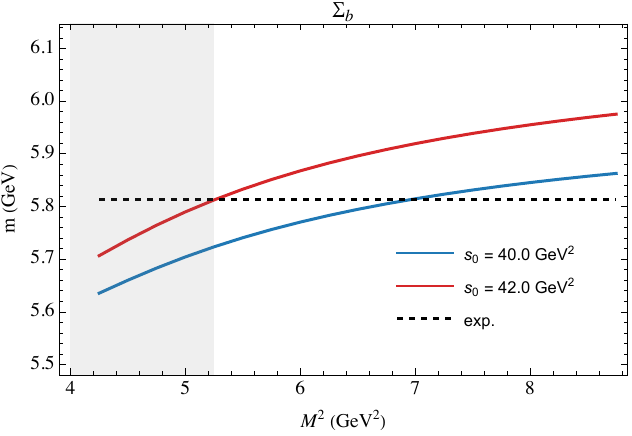}
\caption{Baryon mass from the $\hat p$ mass sum rule ($\beta=-1$) for the lower and upper value of $s_0$. Shaded:
Borel window; dashed: experimental mass.}
\label{fig:mass}
\end{figure*}

\emph{Borel window.} We apply the same criteria to both flavors: the pole contribution to the mass sum rule,
$\int^{s_0}/\int^{\infty}$, is at least $50\%$ at the upper end of the window for all $s_0$ considered, and operators
of dimension $\geq6$ contribute at most $15\%$ of the mass OPE at the lower end. For the Ioffe current this gives
$M^2=1.8$--$2.2$~GeV$^2$ with $s_0=9$--$10$~GeV$^2$ for $\Sigma_c$, and $M^2=4.0$--$5.25$~GeV$^2$ with
$s_0=40$--$42$~GeV$^2$ for $\Sigma_b$. In these windows the pole contribution to the magnetic sum rule exceeds $80\%$.
The mass extracted from the mass sum rule, $m^2=-d\ln\Pi_{\hat p}/d(1/M^2)$, is $2.40$--$2.57$~GeV for $\Sigma_c$ and
$5.64$--$5.81$~GeV for $\Sigma_b$ (Fig.~\ref{fig:mass}), compared with $2.454$ and $5.813$~GeV; it rises across the
window without a plateau, which reflects the simple pole-plus-continuum ansatz. The thresholds
$\sqrt{s_0}\simeq3.0$--$3.2$~GeV and $6.3$--$6.5$~GeV lie above the negative-parity $\Sigma_c(2800)$ and
$\Sigma_b(6097)$, which contribute to the $\hat p$ structures with the same sign as the ground state; separating them
would require parity-projected sum rules, and this effect is not included in our uncertainties.

\emph{OPE structure.} At $\beta=-1$ the $W_{E1}$ OPE in the windows consists of the perturbative term
($54$--$78\%$), dimension-six terms ($24$--$50\%$, mostly $\chi\qqbar^2$, as for the nucleon~\cite{Ioffe:1984}),
dimension-eight terms ($-2$ to $-8\%$) and a gluon-condensate contribution of about $1\%$ (ranges over the windows,
the values of $s_0$ and the states). At the lower end of the windows, where the condensates are largest, the
dimension-six and dimension-eight terms are $40$--$50\%$ and $-4$ to $-8\%$ of the $W_{E1}$ OPE, so that the
expansion converges as well as for the mass sum rule. Our $\kappa$ and $\xi$ terms follow the definitions given with Eq.~(\ref{eq:Sq}) and the values of
Table~\ref{tab:inputs}; the corresponding terms of Ref.~\cite{Zhu:1997} have different coefficients, which
amounts to a different definition of $\xi$. Doubling the $\xi$ term changes the moments by $+0.18$, $-0.09$, $+0.15$
and $-0.08\,\mu_N$ for $\Sigma_c^{++}$, $\Sigma_c^0$, $\Sigma_b^+$ and $\Sigma_b^-$, well within the uncertainties. Through the $\chi\qqbar^2$
term, $\mu$ depends approximately linearly on $\chi$, with the slopes $d\mu/d\chi$ in Table~\ref{tab:results}.
Leaving out all gluon-condensate terms changes $\mu_{\Sigma_c^{++}}$ by about $-0.01\,\mu_N$, and leaving out the
mixed condensate by about $-0.12\,\mu_N$. With $\chi=-4.5$~GeV$^{-2}$, the value used in Ref.~\cite{Zhu:1997}, our sum
rule gives $\mu_{\Sigma_c^{++}}\simeq1.9\,\mu_N$ at central inputs, compared with $2.4\,\mu_N$ for the calibrated $\chi$.

\begin{figure*}[t]
\centering
\includegraphics[width=0.45\textwidth]{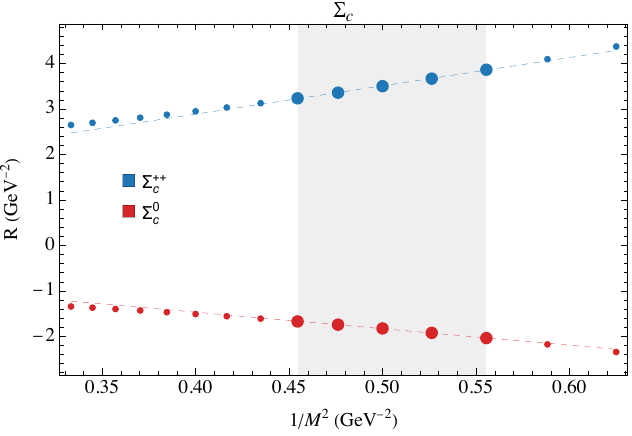}\hfill
\includegraphics[width=0.45\textwidth]{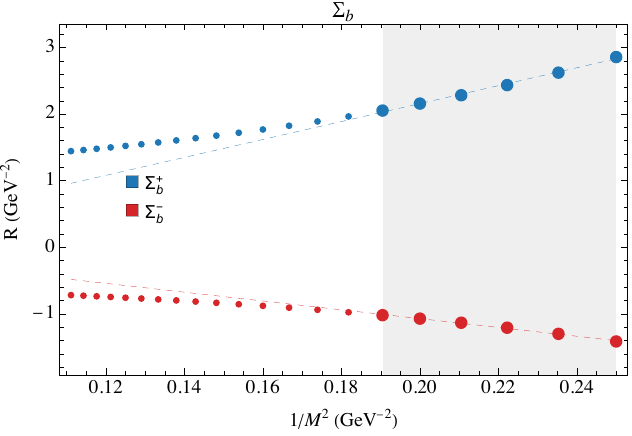}\\[2mm]
\includegraphics[width=0.45\textwidth]{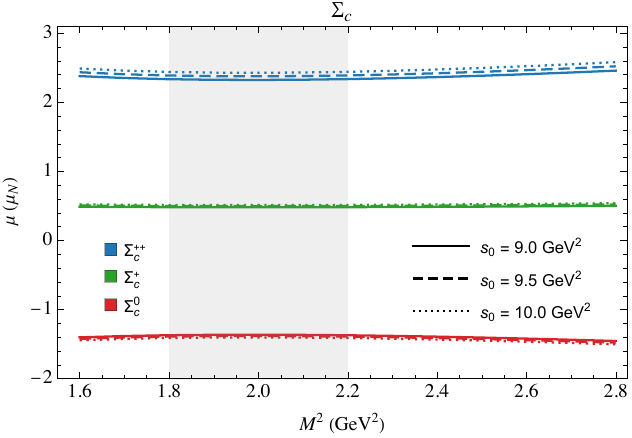}\hfill
\includegraphics[width=0.45\textwidth]{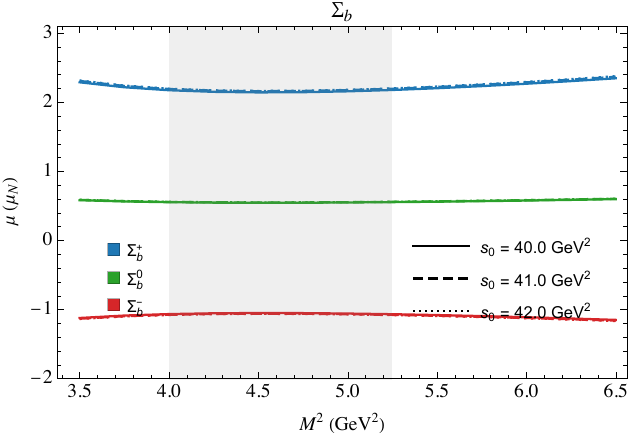}
\caption{Upper panels: the ratio $R$ of Eq.~(\ref{eq:ratio}) at central inputs ($\beta=-1$, $s_0=9.5$ and $41$~GeV$^2$). Large
symbols: points in the Borel window (shaded); dashed: linear fit in the window, whose slope is $\mu$ in natural
magnetons. Lower panels: magnetic moments as functions of $M^2$ at central inputs for three values of $s_0$,
$\mu(M^2)=(m_N/m_B)M^2[R(M^2)-A]$, with $A$ fitted in the window at the same $s_0$.}
\label{fig:ratio}
\end{figure*}

\emph{Ratio.} The upper panels of Fig.~\ref{fig:ratio} show $R$ over a wide range of $M^2$. Inside the windows it is well described by
a straight line, but the lever arm in $1/M^2$ is short ($0.10$ and $0.06$~GeV$^{-2}$), so that the linear form is not
a strong test; at larger $M^2$, where the continuum dominates, $R$ deviates from the line. Correspondingly, the slope
and the intercept are strongly anticorrelated, ${\rm corr}(\mu,A)\simeq-0.7$ ($\Sigma_c$) and $-0.94$ ($\Sigma_b$),
and the result depends on the choice of the window, which we include as an uncertainty.
As a cross-check of the normalization we also use the conventional procedure, with $\tilde\lambda^2$ from the mass
sum rule with the experimental baryon mass, averaged over the window, and a fit of $\Pi_{E1}$ alone. This gives
$2.21$ and $-1.30\,\mu_N$ for $\Sigma_c^{++}$ and $\Sigma_c^0$ (ratio method: $2.38$ and $-1.39$) and $2.70$ and
$-1.34\,\mu_N$ for $\Sigma_b^+$ and $\Sigma_b^-$ (ratio method: $2.18$ and $-1.07$) at central inputs. The differences
reflect the variation of the extracted mass across the windows (Fig.~\ref{fig:mass}); they are largest for $\Sigma_b$,
whose mass sum rule falls below the experimental mass in most of the window. We include half of them as a
normalization uncertainty, $0.26$ and $0.14\,\mu_N$ for $\Sigma_b^+$ and $\Sigma_b^-$ and below $0.1\,\mu_N$ for $\Sigma_c$.
The lower panels show the moments as functions of $M^2$ for three values of $s_0$, defined as
$\mu(M^2)=(m_N/m_B)\,M^2[R(M^2)-A]$ with $A$ from the fit in the window at the same $s_0$. Inside the windows $\mu(M^2)$
varies by less than $1.5\%$, which partly reflects this construction; the dependence on $s_0$ is about $4\%$ for
$\Sigma_c$ and $1\%$ for $\Sigma_b$ across the ranges considered, and outside the windows the curves rise slowly as
the continuum contribution grows.

\emph{Monte Carlo and uncertainties.} We generate 1000 sets of inputs according to Table~\ref{tab:inputs}, with
$\chi$ calibrated on the proton in each set, $s_0$ uniformly distributed in its range and $m_Q$ according to its PDG uncertainty. Each set is used for all six states,
so that correlations between them are kept; the $I_3=0$ moments are averaged sample by sample. For each set we fit
Eq.~(\ref{eq:ratio}) on the Borel grid inside the window and quote the median and the half width of the central $68\%$
interval. We add in quadrature (a)~the interpolating-field uncertainty, half the difference between the results for
$\beta=-0.6$ and $-1.4$, i.e.\ for a $\pm40\%$ variation of the mixing parameter around the Ioffe current with the
same windows and the same calibrated $\chi$. At $\beta=-0.6$ the window criteria of the $\Sigma_Q$ sum rules are satisfied; at
$\beta=-1.4$ they are marginally violated at the edges (pole contribution $45$ and $48\%$ at the upper end of the
windows for the lower $s_0$, dimension $\geq6$ fraction $16\%$ for $\Sigma_c$), which we accept for this estimate;
the pole contribution to the magnetic sum rule stays above $80\%$ throughout; (b)~the window uncertainty, half the spread of the central values for
windows shifted or enlarged by $0.2$~GeV$^2$ ($\Sigma_c$) or $0.5$~GeV$^2$ ($\Sigma_b$) at either end; (c)~the
scale uncertainty of the $\overline{\rm MS}$ mass, the shift of the central value when $\overline m_Q(\overline m_Q)$
is replaced by $\overline m_Q(2\overline m_Q)$, i.e.\ $m_c=1.034$ and $m_b=3.735$~GeV (three-loop running with
$\alpha_s(M_Z)=0.118$), quoted as a symmetric uncertainty although the shift lowers $|\mu|$, while the pole mass
raises it; (d)~the normalization uncertainty, half the difference between the ratio method and the
$\tilde\lambda^2$ procedure described above; and (e)~the calibration uncertainty, $|d\mu/d\chi|\,\delta\chi$
combined with the scale shift of the $\chi\qqbar^2$ term described above. With the calibrated $\chi$ the Monte Carlo uncertainty is three to four times
smaller than for a fixed range $\chi=-6\pm2$~GeV$^{-2}$; the remaining correlation with $\chi$, which now varies with the
other condensates, is $|{\rm corr}(\mu,\chi)|\simeq0.2$--$0.5$. The PDG uncertainty of the $\overline{\rm MS}$ masses contributes
$|d\mu/dm_Q|\,\delta m_Q<0.01\,\mu_N$, much less than their scale dependence. The total uncertainties are comparable to those obtained for the octet moments in the external-field method with
conservative input widths~\cite{Wang:2008oct} and to the accuracy of leading-order sum rules estimated in
Ref.~\cite{Leinweber:1997}; analyses that vary only the QCD inputs quote smaller uncertainties~\cite{Zhu:1997,Ozdem:2025}. The light- and heavy-quark contributions follow from setting $e_Q=0$ or $e_q=0$,
since the OPE is linear in the charges. The results are collected in Table~\ref{tab:results}.

\emph{Heavy-quark mass scheme.} At leading order the scheme of $m_Q$ is not fixed. With the pole masses
$m_c=1.67$ and $m_b=4.78$~GeV and the same windows and thresholds, the moments change by the amounts given in the last
column of Table~\ref{tab:results}, e.g.\ $+0.67\,\mu_N$ for $\Sigma_c^{++}$. However, with pole masses the pole
contribution to the mass sum rule at the upper end of these windows is only $29$--$47\%$, and no choice of $s_0$ satisfies the pole
criterion and reproduces the baryon mass at the same time (the mass sum rule then gives $2.7$ and $6.2$~GeV for
$s_0$ large enough for $50\%$ pole dominance). The $\overline{\rm MS}$ masses lead to a consistent analysis, and we
regard the pole-mass shifts as an upper estimate of the scheme dependence rather than as part of the uncertainty;
the scale variation (c) of the $\overline{\rm MS}$ mass, which moves $m_Q$ in the opposite direction, is included.

\begin{table*}[t]
\caption{Magnetic moments (in $\mu_N$) for $\beta=-1$ with $\overline{\rm MS}$ heavy-quark masses: median and total
uncertainty, with its Monte Carlo (MC), interpolating-field ($\beta$), window, mass-scale, calibration and normalization components; light- and heavy-quark
contributions $\mu_q$, $\mu_Q$; transition strength $A$ (GeV$^{-2}$); sensitivities $d\mu/d\chi$ (in
$\mu_N\,$GeV$^2$) and $d\mu/dm_Q$ (in $\mu_N\,$GeV$^{-1}$) at central inputs; shift of the central value for pole
masses (see text).}
\label{tab:results}
\centering\footnotesize\setlength{\tabcolsep}{2.8pt}
\begin{tabular}{lcccccccccccccc}
\toprule
State & $\mu$ & MC & $\beta$ & window & scale & $\chi$ cal. & norm. & $\mu_q$ & $\mu_Q$ & $A$ & $d\mu/d\chi$ & $d\mu/dm_Q$ & pole mass\\
\midrule
$\Sigma_c^{++}$ & $2.37\pm0.65$ & $0.18$ & $0.41$ & $0.17$ & $0.32$ & $0.29$ & $0.08$ & $2.50\pm0.19$ & $-0.134\pm0.023$ & $0.39\pm0.18$ & $-0.38$ & $1.43$ & $+0.67$\\
$\Sigma_c^{+}$ & $0.50\pm0.15$ & $0.04$ & $0.10$ & $0.03$ & $0.06$ & $0.08$ & $0.02$ & $0.63\pm0.05$ & $-0.134\pm0.023$ & $0.19\pm0.05$ & $-0.10$ & $0.26$ & $+0.12$\\
$\Sigma_c^{0}$ & $-1.39\pm0.37$ & $0.11$ & $0.21$ & $0.11$ & $0.21$ & $0.15$ & $0.04$ & $-1.25\pm0.09$ & $-0.134\pm0.023$ & $-0.02\pm0.10$ & $0.19$ & $-0.92$ & $-0.43$\\
\midrule
$\Sigma_b^{+}$ & $2.16\pm0.65$ & $0.19$ & $0.19$ & $0.26$ & $0.36$ & $0.28$ & $0.26$ & $2.15\pm0.19$ & $0.013\pm0.003$ & $-0.54\pm0.17$ & $-0.26$ & $0.93$ & $+0.80$\\
$\Sigma_b^{0}$ & $0.55\pm0.16$ & $0.05$ & $0.05$ & $0.07$ & $0.09$ & $0.07$ & $0.06$ & $0.54\pm0.05$ & $0.013\pm0.003$ & $-0.14\pm0.04$ & $-0.07$ & $0.24$ & $+0.20$\\
$\Sigma_b^{-}$ & $-1.06\pm0.32$ & $0.09$ & $0.10$ & $0.13$ & $0.17$ & $0.14$ & $0.14$ & $-1.08\pm0.09$ & $0.013\pm0.003$ & $0.25\pm0.08$ & $0.13$ & $-0.45$ & $-0.39$\\
\bottomrule
\end{tabular}
\end{table*}

\begin{table*}[t]
\caption{Magnetic moments (in $\mu_N$) compared with other approaches. Lattice values: linear/quadratic chiral
extrapolation. Quark models, range: range of the values compiled in Ref.~\cite{Can:2014}.
$^a$In the chiral quark-soliton model the $\Sigma_b$ moments equal those of the $\Sigma_c$ states with the same
light quarks~\cite{Yang:2018}. $^b$Used as input to fix low-energy constants. $^c$Ref.~\cite{Zhu:1997}
gives $0.6\pm0.1$ in units of $e/2m_{\Sigma_c}$, i.e.\ $0.23\pm0.03\,\mu_N$; the abstract quotes the former number with
the unit $\mu_N$. $^d$Simple nonrelativistic quark model as evaluated in Ref.~\cite{Bernotas:2012}. $^e$Refs.~\cite{Mohan:2022}
($\Sigma_c$) and \cite{Dhir:2013} ($\Sigma_b$).}
\label{tab:compare}
\centering\footnotesize
\begin{tabular}{lcccccc}
\toprule
Approach & $\Sigma_c^{++}$ & $\Sigma_c^{+}$ & $\Sigma_c^{0}$ & $\Sigma_b^{+}$ & $\Sigma_b^{0}$ & $\Sigma_b^{-}$\\
\midrule
This work & $2.37\pm0.65$ & $0.50\pm0.15$ & $-1.39\pm0.37$ & $2.16\pm0.65$ & $0.55\pm0.16$ & $-1.06\pm0.32$\\
Light-cone QCDSR~\cite{Ozdem:2025} & $2.02(18)$ & $0.50(5)$ & $-1.01(9)$ & $2.02(19)$ & $0.53(6)$ & $-1.01(9)$\\
External-field QCDSR~\cite{Zhu:1997} & $2.1\pm0.3$ & $0.23\pm0.03^c$ & $-1.6\pm0.2$ & -- & -- & --\\
Lattice QCD~\cite{Can:2014} & $1.57(25)$/$2.22(51)$ & -- & $-0.85(13)$/$-1.07(27)$ & -- & -- & --\\
Chiral quark-soliton model~\cite{Yang:2018} & $2.15(10)$ & $0.46(3)$ & $-1.24(5)$ & $2.15(10)^a$ & $0.46(3)^a$ & $-1.24(5)^a$\\
HB ChPT, $\mathcal O(p^3)$~\cite{Wang:2018} & $1.50^b$ & $0.26(7)$ & $-0.97(5)$ & -- & -- & --\\
Covariant ChPT~\cite{Shi:2019} & $1.285$ & $0.190$ & $-0.905$ & -- & -- & --\\
Relativistic three-quark model~\cite{Faessler:2006} & $1.76$ & $0.36$ & $-1.04$ & $2.07$ & $0.53$ & $-1.01$\\
Nonrelativistic quark model$^d$ & $2.35$ & $0.49$ & $-1.37$ & $2.50$ & $0.64$ & $-1.22$\\
Bag model~\cite{Bernotas:2012} & $1.68$ & $0.32$ & $-1.04$ & $1.62$ & $0.42$ & $-0.78$\\
Bag model~\cite{Simonis:2018} & $2.28$ & $0.49$ & $-1.31$ & $2.25$ & $0.60$ & $-1.15$\\
Eff.\ mass, screened charge$^e$ & $2.10$ & $0.43$ & $-1.23$ & $2.19$ & $0.56$ & $-1.06$\\
Hypercentral model~\cite{Majethiya:2008} & $1.95$ & $0.39$ & $-1.17$ & $2.12$ & $0.55$ & $-1.03$\\
Quark models, range~\cite{JuliaDiaz:2004,Faessler:2006,Bernotas:2012,Sharma:2010,Barik:1983,Kumar:2005,Patel:2008}
 & $1.68$--$3.07$ & -- & $-1.78$ to $-1.02$ & -- & -- & --\\
\bottomrule
\end{tabular}
\end{table*}

\begin{figure}[!tb]
\centering
\includegraphics[width=0.95\columnwidth]{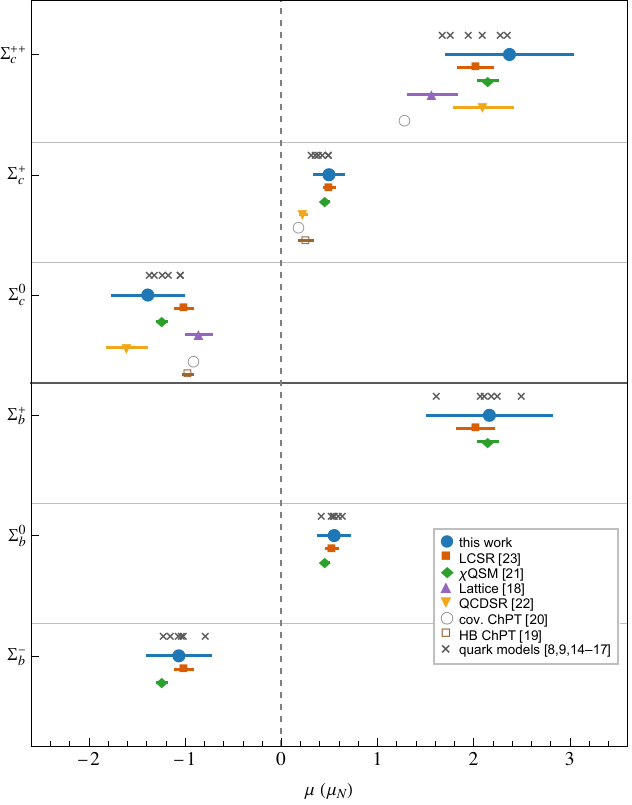}
\caption{Our results (circles, total uncertainty) compared with the chiral quark-soliton model~\cite{Yang:2018}
(for $\Sigma_b$ its moments equal those of the $\Sigma_c$ states with the same light quarks), lattice QCD (linear chiral
fit)~\cite{Can:2014}, external-field~\cite{Zhu:1997} and light-cone~\cite{Ozdem:2025} sum rules, covariant~\cite{Shi:2019}
and heavy-baryon~\cite{Wang:2018} chiral perturbation theory, and quark models (crosses: the six models in
Table~\ref{tab:compare}).}
\label{fig:compare}
\end{figure}

\section{Discussion}

Table~\ref{tab:compare} and Fig.~\ref{fig:compare} compare our results with other approaches. Within the uncertainties they agree with the
chiral quark-soliton model~\cite{Yang:2018} for all six states and, for $\Sigma_c^{++}$ and $\Sigma_c^0$, with the
earlier external-field sum rules~\cite{Zhu:1997}; for $\Sigma_c^+$ the latter give $0.23(3)\,\mu_N$, about $1.8$
standard deviations below our value. The lattice values for $\Sigma_c^{++}$ and $\Sigma_c^0$ are $1.57(25)$ and $-0.85(13)\,\mu_N$
(linear) or $2.22(51)$ and $-1.07(27)\,\mu_N$ (quadratic chiral fit)~\cite{Can:2014}; our results agree with the
quadratic fit and lie $1.2$ and $1.4$ standard deviations above the linear fit in magnitude. Chiral perturbation
theory~\cite{Wang:2018,Shi:2019} gives smaller magnitudes. The agreement with Ref.~\cite{Zhu:1997} for
$\Sigma_c^{++}$ is partly accidental: with their value of $\chi$ our sum rule gives a smaller moment, while the
normalization through the mass sum rule acts in the opposite direction.

The recent light-cone sum-rule analysis~\cite{Ozdem:2025} agrees with our results within about one standard
deviation for all six states; the largest differences, $0.35$ and $0.38\,\mu_N$, occur for $\Sigma_c^{++}$ and
$\Sigma_c^0$, where our magnitudes are larger. The two approaches are complementary: the light-cone sum rules
parametrize the long-distance photon emission through photon distribution amplitudes, whereas in the external-field
method it enters through the susceptibilities $\chi$, $\kappa$ and $\xi$, of which we fix the dominant one, $\chi$,
with the proton moment.
Ref.~\cite{Ozdem:2025} finds nearly equal moments for corresponding $\Sigma_c$ and $\Sigma_b$ states; our central values
for $\Sigma_c^{++}$ and $\Sigma_c^0$ are about $10$--$30\%$ larger in magnitude than those for $\Sigma_b^+$ and
$\Sigma_b^-$, but within our uncertainties the two flavors are also compatible with equal moments (see below). The quark models in Table~\ref{tab:compare} give
$1.7$--$2.4\,\mu_N$ for $\Sigma_c^{++}$ and $1.6$--$2.5\,\mu_N$ for $\Sigma_b^+$; their spread, which reflects different
quark masses, wave functions and treatments of relativistic and screening effects, is comparable to our uncertainty,
and our central values lie at or slightly above its upper end for $\Sigma_c$ and inside it for $\Sigma_b$. All
approaches satisfy, exactly or approximately, the isospin relation $\mu_{\Sigma_Q^{I_3=0}}\simeq(\mu_{\Sigma_Q^{I_3=+1}}+
\mu_{\Sigma_Q^{I_3=-1}})/2$, which holds by construction in our analysis; since the charged moments largely cancel in this
average, the $I_3=0$ moments test their relative size, and our $\Sigma_c^+$ and $\Sigma_b^0$ values,
$0.50\pm0.15$ and $0.55\pm0.16\,\mu_N$, agree well with the light-cone sum rules and most quark models.

The moments are carried by the light diquark. The heavy-quark contribution is $-0.13\pm0.02\,\mu_N$ in $\Sigma_c$ and
$+0.013\pm0.003\,\mu_N$ in $\Sigma_b$, and varies by about $0.01\,\mu_N$ between $\beta=-0.6$ and $-1.4$. At
$\beta=-1$ it comes from mass-dependent condensate terms (Eq.~(\ref{eq:rho})), so that its separation from the
light-quark part depends on the interpolating field and the Dirac structure. The nonrelativistic quark-model terms
$-\mu_Q/3$ are $-0.12$ to $-0.16\,\mu_N$ for charm and about $+0.02\,\mu_N$ for bottom (for quark masses of
$1.3$--$1.7$ and $4.7$--$5$~GeV), of the same sign and size. Since in this model
$\mu_{\Lambda_c^+}=\mu_c$, the planned measurement of the $\Lambda_c^+$ moment with an uncertainty of about
$0.01\,\mu_N$~\cite{ALADDIN:2024} would fix the charm term $-\mu_c/3$ in $\Sigma_c$ to a few thousandths of $\mu_N$.
Taken at face value, our heavy-quark contribution corresponds to $\mu_c\simeq0.40\pm0.07\,\mu_N$, at the upper end
of the range $0.15$--$0.40\,\mu_N$ of the predictions for $\mu_{\Lambda_c^+}$ compiled in Ref.~\cite{ALADDIN:2024};
given the dependence of the separation on the interpolating field noted above, this comparison is qualitative. The lattice calculation finds a charm contribution
$e_c\mu_Q=-0.066(14)\,\mu_N$ (linear) or $-0.039(24)\,\mu_N$ (quadratic fit), and a light-quark contribution
$e_u\mu_q=1.58(24)\,\mu_N$ for $\Sigma_c^{++}$~\cite{Can:2014}, compared with our $2.50\pm0.19\,\mu_N$. Our heavy-quark contribution is thus larger in magnitude than the lattice
value by a factor of two to three ($2.5\sigma$ for the linear fit). Part of this difference may come from the
separation itself: on the lattice it is defined through the electromagnetic current, here through the charge
factors of the OPE of a particular current and Dirac structure at leading order, in which the heavy-quark term is
generated by mass-dependent condensate contributions; the two definitions need not coincide. In the total moment the
difference resides mainly in the light-quark sector.
The light-diquark contributions in $\Sigma_c$ and $\Sigma_b$ differ by
$\mu_q(\Sigma_c^{++})-\mu_q(\Sigma_b^+)=0.35\pm0.41\,\mu_N$, where the uncertainty combines the correlated Monte Carlo
error ($0.06$) with the interpolating-field ($0.22$) and window ($0.34$) systematics; the scale of $m_Q$ cancels in
the difference, and adding the normalization uncertainties of the two states would increase the error to about
$0.49\,\mu_N$. The two flavors are therefore compatible with the heavy-quark limit; the windows lie at different
Borel scales, and anomalous-dimension effects, which we neglect, act between them, so that a deviation of the
expected size $\Lambda_{\rm QCD}/m_c$ cannot be resolved at this order.

The main limitations are the absence of $\mathcal O(\alpha_s)$ and anomalous-dimension corrections, the
heavy-quark mass scheme at leading order, the use of an effective $\chi$ calibrated on the proton, the truncation of
the OPE, and the description of excited states through a single transition constant and a continuum threshold,
including the negative-parity states. With $\chi$ fixed, the interpolating field and the scale of the heavy-quark
mass give the largest remaining uncertainties for $\Sigma_c$, and the scale, the calibration, the window and the
normalization for $\Sigma_b$; next-to-leading-order corrections would reduce the scale dependence. With the slopes in Table~\ref{tab:results} the results can be rescaled to other values of $\chi$.

\FloatBarrier
\section{Summary}

We have computed the magnetic moments of the $\Sigma_c$ and $\Sigma_b$ isotriplets in external-field QCD sum rules
with the full heavy-quark mass dependence, including the couplings of the heavy quark to the photon and to gluons,
and with the mass sum rule of the same current as normalization. The construction reproduces known heavy--light and
light-baryon sum rules, and the mass sum rules give the baryon masses within a few percent. The results in
Table~\ref{tab:results} agree with the chiral quark-soliton model and with the quadratic chiral extrapolation of
lattice QCD for $\Sigma_c$, and with recent light-cone sum rules for all six states; the moments are carried by the
light diquark, and within the uncertainties the $\Sigma_c$ and $\Sigma_b$ moments are compatible with the
heavy-quark limit. Fixing $\chi$ with the proton magnetic moment, which also reproduces the neutron moment, removes the
largest source of uncertainty; the dependence on the heavy-quark mass scheme is the main open issue at this order.  

\appendix
\section{OPE for $\beta=-1$}
\label{app:ope}

For $\beta=-1$, in the normalization of Eq.~(\ref{eq:phen}), with $r=m_Q^2/s$, $E\equiv e^{-m_Q^2/M^2}$,
$E_y\equiv e^{-m_Q^2/((1-y)M^2)}$, $y_0=1-m_Q^2/s_0$, $\langle\bar qg_s\sigma\!\cdot\!Gq\rangle=-m_0^2\qqbar$,
$e_q$ the charge of the two light quarks and $e_Q$ that of the heavy quark,
\begin{align}
\Pi_{i}&=\int_{m_Q^2}^{s_0}\!ds\,e^{-s/M^2}\rho_{i}(s)+\Pi^{\rm np}_{i},\qquad i=\hat p,\,E1,\nonumber\\
\rho_{\hat p}&=\frac{s^2}{64\pi^4}\left(1-8r+8r^3-r^4-12r^2\ln r\right)\nonumber\\[3pt]
&\quad+\frac{\GG}{128\pi^4}(1-r)^2,\nonumber\\
\rho_{E1}&=-\frac{e_q}{32\pi^4}\,s(1-r)^3,\nonumber\\
\Pi^{\rm np}_{\hat p}&=-\frac{m_Q^2\GG}{192\pi^4}\int_0^{y_0}\!dy\,\frac{y^2}{(1-y)^2}E_y\nonumber\\
&\quad+E\Big[\frac23\qqbar^2\nonumber\\
&\qquad+\frac{M^2+2m_Q^2}{6M^4}\qqbar\mix\Big],\nonumber\\
\Pi^{\rm np}_{E1}&=\frac{\GG}{2304\pi^4}\int_0^{y_0}\!dy\Big[\frac{6e_qm_Q^2\,y}{(1-y)^2M^2}\nonumber\\
&\qquad-\frac{e_Q(3-5y)}{1-y}-12e_q\Big]E_y\nonumber\\
&\quad+\frac{e_Q\GG^2}{165888\,\pi^4M^4}\nonumber\\
&\qquad\times\int_0^{y_0}\!dy\Big[\frac{3m_Q^2}{(1-y)^3M^2}-\frac{2}{(1-y)^2}\Big]E_y\nonumber\\
&\quad+E\qqbar^2\Big[\frac{3e_QM^2+2e_q(M^2+m_Q^2)}{18M^4}+\frac{e_q\chi}{3}\nonumber\\
&\qquad+\frac{e_q\kappa(m_Q^2-2M^2)+e_q\xi(M^2+m_Q^2)}{18M^4}\Big]\nonumber\\
&\quad+E\qqbar\mix\Big[\frac{e_q\chi(M^2+2m_Q^2)}{24M^4}\nonumber\\
&\qquad-\frac{2e_QM^2(M^2-3m_Q^2)+e_qm_Q^2(M^2-2m_Q^2)}{72M^8}\Big].
\end{align}
The terms with $E$ are pole terms at $s=m_Q^2$; the $y$ integrals contain the continuum-subtracted gluon-condensate
contributions. The perturbative heavy-quark term and all terms linear in $\qqbar$ are proportional to $1+\beta$ and
vanish here; they are included for $\beta\neq-1$.

\end{document}